\documentclass[pre, 10pt, showkeys]{revtex4-2}

\usepackage{etex} 

\usepackage{amsmath, amssymb, braket, commath, bm, empheq, array} 

\usepackage{subfiles}

\usepackage[]{graphicx}
\usepackage{hyperref}
\hypersetup{
	colorlinks=true,
	linkcolor=red,
	filecolor=blue,
	urlcolor=magenta,
}

\usepackage[caption=false, font=footnotesize, subrefformat=parens, labelformat=parens, belowskip=-0pt, aboveskip=-0pt,position=top]{subfig}

\usepackage{dblfloatfix}

\usepackage{booktabs, multirow}

\renewcommand{\thesection}{\arabic{section}} 

\newcommand{\bte}{$\beta$-ensemble}

\newcommand{\fbslk}[2]{\ensuremath{\text{K}_{#1}\del{#2}}}

\newcommand{\fnrm}[1]{\ensuremath{||#1||_\text{F}}}	 
\newcommand{\gat}{\ensuremath{\gamma_{\text{AT}}}}	
\newcommand{\get}{\ensuremath{\gamma_{\text{ET}}}}	
\newcommand{\gtat}{\ensuremath{\tilde{\gamma}_{\text{AT}}}}	
\newcommand{\gtet}{\ensuremath{\tilde{\gamma}_{\text{ET}}}}	
\newcommand{\prob}[1]{\ensuremath{\text{P}\del{#1}}}
\newcommand{\mean}[1]{\ensuremath{\left\langle#1\right\rangle}}

\newcommand{\ra}{\ensuremath{\mathcal{R}_1}}
\newcommand{\rb}{\ensuremath{\mathcal{R}_2}}
\newcommand{\shn}{\ensuremath{\textrm{S}}}
\newcommand{\tTh}{\ensuremath{t_{\textrm{Th}}}}		
\newcommand{\tR}{\ensuremath{t_{\textrm{R}}}}		

\newcommand{\xxx}{1-D disordered spin-1/2 Heisenberg model}
\begin{document}
\title{Non-ergodic extended states in the \bte}
\author{Adway Kumar Das}\email{akd19rs062@iiserkol.ac.in}
\author{Anandamohan Ghosh}\email{anandamohan@iiserkol.ac.in}
\affiliation{
	Indian Institute of Science Education and Research Kolkata, Mohanpur, 741246 India\\
	Department of Physical Sciences
}

\date{\today}
\begin{abstract}
Matrix models showing chaotic-integrable transition in the spectral statistics are important for understanding Many Body Localization (MBL) in physical systems. One such example is the \bte, known for its structural simplicity. However, eigenvector properties of \bte\ remain largely unexplored, despite energy level correlations being thoroughly studied. In this work we numerically study the eigenvector properties of \bte\ and find that the Anderson transition occurs at $\gamma = 1$  and ergodicity breaks down at $\gamma = 0$ if we express the repulsion parameter as $\beta = N^{-\gamma}$. Thus other than Rosenzweig-Porter ensemble (RPE), \bte\ is another example where Non-Ergodic Extended (NEE) states are observed over a finite interval of parameter values ($0<\gamma<1$).
We find that the chaotic-integrable transition coincides with the breaking of ergodicity in \bte\ but with the localization transition in the RPE or the \xxx. As a result, the dynamical time-scales in the NEE regime of \bte\ behave differently than the later models.
\end{abstract}
\maketitle
\section{Introduction}\label{sec_intro}
Canonically invariant classical ensembles including Dyson's threefold ways \cite{Dyson1} and their extensions over symmetric spaces \cite{Zirnbauer1, Ivanov1} (e.g. Laguerre \cite{Dubbs1}, Jacobi \cite{Dubbs2}, Circular \cite{Killip1} ensembles) are central to the paradigm of Random Matrix Theory (RMT) \cite{Mehta1} epitomizing completely ergodic \cite{Borgonovi1} and chaotic \cite{Bohigas1} dynamics in quantum mechanical systems. Corresponding energy levels tend to repel each other, where the degree of repulsion is called the Dyson's index, $\beta$ having values 1, 2 and 4 for the Gaussian Orthogonal, Unitary and Symplectic ensembles respectively \cite{Guhr1}. On the other hand, regular dynamics observed in integrable systems \cite{Yuzbashyan1, Corrigan1} is usually captured by the Poisson ensemble \cite{Berry3}, where energy levels are uncorrelated with inclination to be clustered (hence can be assigned $\beta = 0$). However, several physical systems (e.g. kicked top \cite{Haake1}, pseudo-integrable billiards \cite{Abul-Magd1}, Harper \cite{Evangelou1}, Anderson model \cite{Shklovskii1} etc.) show a spectral property intermediate between the aforementioned ideal limits. While phenomenological models \cite{Brody1, Berry2, Izrailev1} can mimic the spectral properties in the intermediate regions, there exist several generalizations of the classical ensembles capturing the physics of mixed dynamics \cite{Rosenzweig1, Carvalho1, Casati1, Mirlin2, Toscano1, Abul-Magd3, Canali1, Pandey1, Bantay1}. In particular, the Joint Probability Distribution Function (JPDF) of eigenvalues for the classical ensembles can be expressed as a Gibbs-Boltzmann weight of a 2-D system of particles, known as the Coulomb gas model \cite{Dyson2}, where $\beta$ is no longer restrained to be quantized. Specifically, a harmonic confining potential yields the Gaussian (also known as the Hermite) \bte\ characterized by the following JPDF,
\begin{align}
	\label{eq_beta_jpdf}
	\begin{split}
		\prob{\vec{E}} &= \frac{1}{\mathcal{Z}_\beta} \exp\del{-\sum_{i=1}^{N} \dfrac{E_i^2}{2}}\prod_{i<j}\abs{E_i - E_j}^\beta
	\end{split}
\end{align}
where $\mathcal{Z}_\beta$ is the normalization constant and $\vec{E} = \{E_1, E_2, \dots, E_N\}$ is the set of $N$ eigenvalues \cite{Forrester2}. Such ensembles were originally conceived as lattice gas systems \cite{Baker1} in connection to the ground state wave-functions of the Calogero-Sutherland model \cite{Choquard1}. Following the rescaling $E_i\to \sqrt{\beta N}E_i$, we can express the partition function $\mathcal{Z}_\beta$ as \cite{Livan1},
\begin{align}
	\label{eq_beta_Z}
	\begin{split}
		\mathcal{Z}_\beta &\propto \int_{\mathbb{R}^N} \prod_{j = 1}^{N} dE_j\exp\del{-\beta N^2 \mathcal{V}[\vec{E}]},\quad \mathcal{V}[\vec{E}] = \sum_{i = 1}^{N}\dfrac{E_i^2}{2N} - \sum_{i\neq j}\dfrac{\log\abs{E_i - E_j}}{2N^2}
	\end{split}
\end{align}
where the potential $\mathcal{V}[\vec{E}]$ has a confining term competing with the pairwise logarithmic repulsion between $N$ fictitious particles. The strength of such interactions is controlled by $\beta$ \cite{Caer1}, which can be interpreted as the inverse temperature. In the infinite temperature limit ($\beta\to 0$), the energy levels are allowed to come arbitrarily close to each other, resulting in Poisson statistics, i.e. a signature of integrability \cite{Berry3}. On the other hand, for $\beta = 1$, Eq.~\eqref{eq_beta_jpdf} coincides with the JPDF of Gaussian Orthogonal Ensemble (GOE), yielding Wigner-Dyson statistics characterized by complete level repulsion, i.e. a signature of chaos \cite{Bohigas1}. Thus tuning $\beta$, it is possible to control the degree of level repulsion in the energy spectrum of \bte\ with $\beta = 1$ indicating the chaotic-integrable transition. Corresponding Hamiltonians can be represented as real, symmetric and tridiagonal $N\times N$ matrices, $H$, with following non-zero elements \cite{Dumitriu1},
\begin{align}
	\label{eq_H_elements}
	H_{i,i} = A_i,\: H_{i, i+1} = H_{i+1, i} = B_i/\sqrt{2}, \quad A_i\sim \mathcal{N}(0, 1),\: B_i\sim \chi_{(N-i)\beta}
\end{align}
where $\mathcal{N}(0, 1)$ is the Normal distribution and $\chi_k$ is the Chi-distribution with a degree of freedom $k$. There has been extensive studies on \bte\ in terms of the Density of States (DOS) \cite{Pandey4, Baker1, Dumitriu4, Albrecht1}, associated fluctuations \cite{Johansson1, Desrosiers2, Killip2, Forrester7, Monvel1}, connection to stochastic differential operators \cite{Valko1, Edelman2, Ramirez1}, extreme eigenvalues \cite{Dumaz1, Borot1, Allez1, Edelman4}. 
In this work we give numerical evidence of chaotic$\rightarrow$integrable, ergodic$\rightarrow$non-ergodic and delocalization$\rightarrow$localization transitions by thoroughly studying the properties of eigenvalues (Sec.~\ref{sec_energy}), eigenfunctions (Sec.~\ref{sec_state}) and dynamics (Sec.~\ref{sec_dynamics}). Therefore we identify the critical values of $\beta$ segregating ergodic, Non-Ergodic Extended (NEE) and localized regimes. We compare the spectral properties of the \bte\ with the  properties of another matrix model, namely Rosenzweig-Porter ensemble (RPE) \cite{Rosenzweig1, Das1}, where for a real symmetric matrix, $H$, all the elements are randomly distributed with
\begin{align}
	\label{eq_H_rpe}
	H_{i,i}\sim \mathcal{N}(0, 1),\: H_{i,j}/\sigma \sim \mathcal{N}(0, 1),\quad \sigma^2 = 1/2N^{\tilde{\gamma}},\: \tilde{\gamma}\in \mathbb{R}
\end{align}	
Moreover, as an impetus to applications in physical systems, we compare both \bte\ and RPE to the widely studied \xxx\ \cite{Santos2, Torres2, Luitz1}, where for simplicity we will consider the chain to be isotropic. The Hamiltonian of such a chain of length $L$ with magnetic field in the Z-direction is defined as,
\begin{align}
	\label{eq_xxx_H}
	H &= -\dfrac{J}{2}\sum_{i=1}^{L}\vec{\sigma_i}\cdot\vec{\sigma}_{i+1} + h_i\hat{\sigma_i}^z
\end{align}
where $\vec{\sigma_i} = \cbr{\hat{\sigma_i}^x, \hat{\sigma_i}^y, \hat{\sigma_i}^z}$ are the Pauli matrices, $h_i$ is the random magnetic field applied in Z-direction on the $i^{th}$ site and $J$ is the coupling constant. We assume periodic boundary condition (i.e. $\hat{\sigma}_{i+L}^\alpha = \hat{\sigma}_i^\alpha$), $J = 1$ and $h_i$ to be uniform random numbers sampled from $[-h, h]$, i.e. the disorder in Heisenberg model can be controlled by tuning $h$. While the model is integrable exactly at $h = 0$, even an infinitesimal fluctuation in magnetic fields on different sites is expected to induce chaos in the thermodynamic limit ($L\to \infty$) \cite{Torres2}. Increasing $h$ further introduces more defects in the chain leading to Many-Body Localization (MBL) where the critical disorder strength, $h_c$ for ergodic to MBL transition depends on the energy density 
\cite{Luitz1}. Since the Z-component of total spin, $S_z = \dfrac{1}{2}\sum_{i = 1}^{L}\hat{\sigma}_i^z$ is conserved in the Heisenberg model, we take $L$ to be even and choose the largest symmetry sector $S_z = 0$ having $L\choose L/2$ eigenvalues for our analysis.

One can intuitively speculate the existence of two critical points in \bte, considering that the diagonal part, $A$, is competing with the perturbation from the off-diagonal part $B$. 
The overall interaction strength can be calculated in terms of the Frobenius norm of $B$, i.e. $\fnrm{B} \approx \sqrt{\sum_{i = 1}^{N-1}\overline{B_i^2}} = \sqrt{\frac{1}{2}N(N-1)\beta} \approx N\sqrt{\beta}$, where $\overline{B_i^2} = (N-i)\beta$ is the mean value of $B_i^2$ (as $\chi^2_k$ distribution has mean $k$). Similarly, the strength of diagonal contribution is $\fnrm{A} \approx \sqrt{N}$.
Thus for weak perturbation ($\fnrm{B}<\fnrm{A}\Rightarrow \beta<1/N$), one may expect the energy states of \bte\ to localize while they should be extended for $\fnrm{B}>\fnrm{A}$. Therefore it is convenient to express the Dyson's index as
\begin{align}
	\label{eq_beta_gamma}
	\beta = N^{-\gamma},\quad \gamma\in\mathbb{R}.
\end{align}
So the perturbation strength can be expressed as $\fnrm{B} = N^{1-\gamma/2}$ and it is reasonable to expect that $\gat \equiv 1$ is the Anderson transition point such that the energy states are localized for $\gamma > \gat$.

The role of the control parameter manifested in the off-diagonal terms of \bte\ is reminiscent of the RPE, where relative strength of perturbation indicates that Anderson transition occurs at $\gtat = 2$ \cite{Kravtsov1}. Moreover, due to the random sign altering nature of the RPE matrix elements, there exists an ergodic transition at $\gtet = 1$ segregating three distinct phases: ergodic, NEE and localized states \cite{Kravtsov1}. 
Similarly, for \bte, even though the off-diagonal elements, $B_i$'s are strictly positive, the diagonal terms, $A_j$'s can be positive or negative at random. If we equate the rescaled perturbation, $\fnrm{B}/\sqrt{N}$ to the total fluctuation from on-site terms, $\fnrm{A}$, we expect ergodic transition at $\get \equiv 0$ such that the energy states occupy the entire Hilbert space in the regime $\gamma \leq \get$. However, these heuristic arguments based on the norm alone cannot account for any phase transition. For example, eigenvectors are exponentially localized in tridiagonal matrices with i.i.d. random elements \cite{Schenker1}. Thus the inhomogeneity of the off-diagonal terms evident from Eq.~\eqref{eq_H_elements} is essential for the existence of NEE in the \bte\ as will be demonstrated in the following sections.
\section{Properties of Energy Levels}\label{sec_energy}
\begin{figure}[t]
	\centering
	\includegraphics[width=\textwidth, trim = {0 375 0 380}]{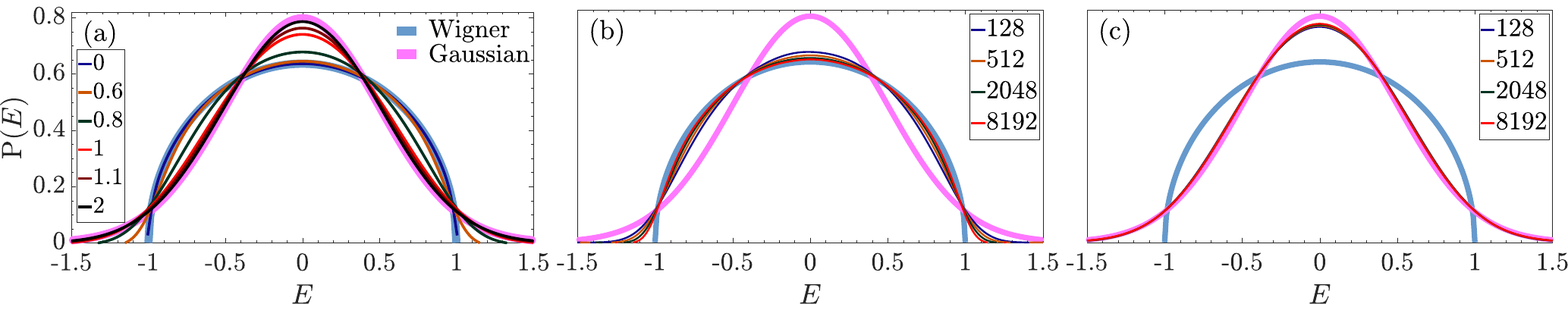}
	\caption{\textbf{Density of States (DOS)} of \bte\	
		(a) averaged over 500 disordered realizations various $\gamma$ where $N = 8192$. (b) $\gamma = 0.6$ and (c) $\gamma = 1.1$ for various $N$. Wigner's semicircle law and Gaussian distribution $\mathcal{N}(0, 1/4)$ shown via bold curves.
	}
	\label{fig_1}
\end{figure}

Now we would like to study the energy level properties of \bte\ following the Hamiltonian in Eq.~\eqref{eq_H_elements} and identify the transition from integrable to chaotic regimes as we vary the Dyson's index. We also compare the properties of \bte\ with the results known for RPE and Heisenberg model. Some of the results in this section are known and we list them for completeness. 
\paragraph{\textbf{Density of States (DOS):}} As a starting point we look at the DOS, which is the marginal PDF of energy levels. The bulk eigenvalues of \bte\ roughly scale with system size as $\epsilon_\beta = \sqrt{4+2N^{1-\gamma}}$ \cite{Dumitriu2}. So we scale the eigenvalues as $E\to E/\epsilon_\beta$ and obtain the DOS numerically, as shown in Fig.~\ref{fig_1}(a) for $N = 8192$ and various $\gamma$. For $\gamma>1$ we obtain $\epsilon_\beta\approx 2$ such that the DOS converges to the Gaussian distribution, $\mathcal{N}(0, 1/4)$ with increasing $N$, as illustrated in Fig.~\ref{fig_1}(c) for a specfic value of $\gamma = 1.1$. Exactly at $\gamma = 1$, the DOS is system size independent and follows a shape intermediate between Gaussian distribution and Wigner semi-circle law since $\epsilon_\beta = \sqrt{6}$. But for $0\leq \gamma<1$, the DOS converges to the semicircle law upon increasing $N$, as shown in Fig.~\ref{fig_1}(b) for a specific value of $\gamma = 0.6$. However, in the limit $\beta\to \infty$ (i.e. $\gamma<0$ and $N\to\infty$), all the eigenvalues of \bte\ freeze and produce a picket-fence spectrum \cite{Gustavsson1}. A qualitative evolution of the shape of the DOS in $\beta$-$N$ plane is shown in Fig.~2 of \cite{Caer1}.

\begin{figure}[t]
	\centering
	\includegraphics[width=\textwidth, trim = {0 260 0 265}]{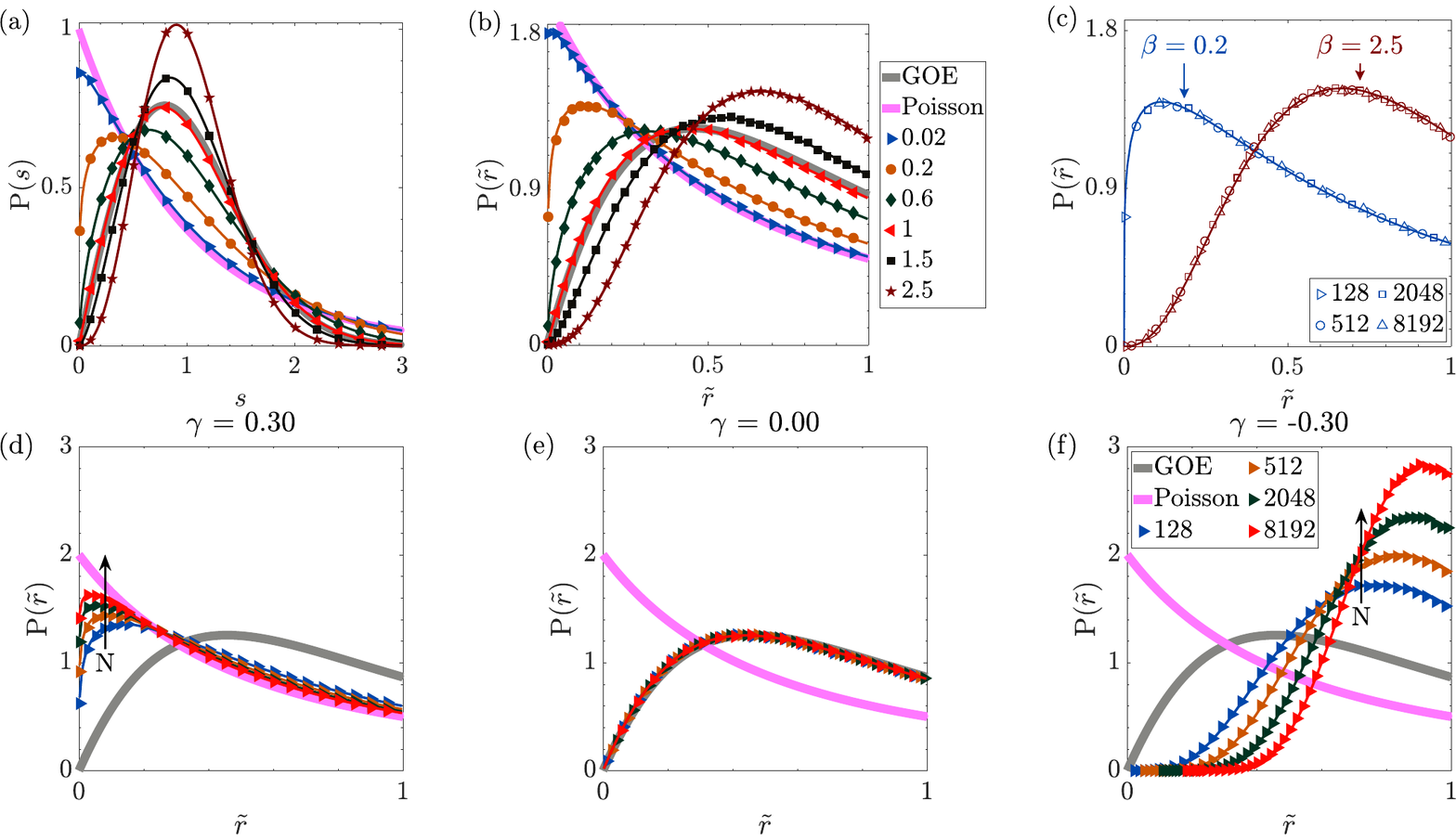}
	\caption{\textbf{Short range spectral correlations} for \bte: PDF of (a)~{NNS} and {modified RNNS} (b) varying $\beta$ for $N = 8192$, varying $N$ (c) for $\beta = 0.2$ (blue markers) and $\beta = 2.5$ (maroon markers) (d) $\gamma = 0.3$ (e) $\gamma = 0$ and (f) $\gamma = -0.3$. The markers indicate numerical data while solid lines denote empirical analytical forms (Eqs.~(12) and (13) in \cite{Caer1} for NNS and Eq.~(1) in \cite{Corps1} for RNNS). The analytical expressions for Poisson and GOE are shown via bold curves.
	}
	\label{fig_2}
\end{figure}

\begin{figure}[t]
	\centering
	\includegraphics[width=\textwidth, trim = {0 275 0 275}]{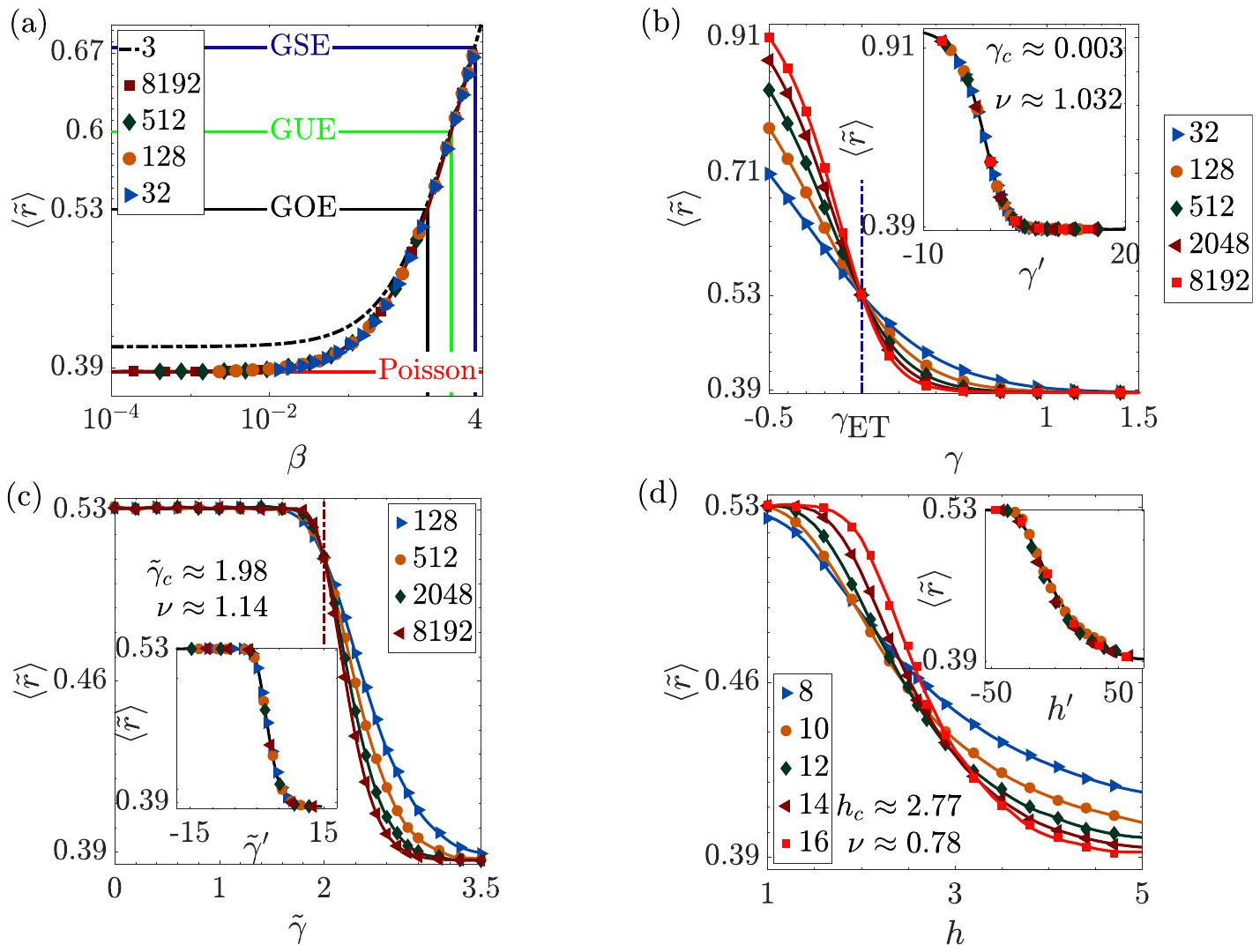}
	\caption{\textbf{Ensemble average of $\tilde{r}$} for \bte\ vs. (a) $\beta$ (b) $\gamma$ (c) for RPE vs. $\tilde{\gamma}$ and (d) Heisenberg model vs. disorder strength, $h$. For (a), (b), (c) system size, $N$ and for (d) chain length, $L$ is varied. In (a) we show the $\mean{\tilde{r}}$ for $N = 3$ (Eq.~(7) in \cite{Atas1}) via dashed line. The insets of (b), (c) and (d) show collapsed data following the ansatz in Eq.~\eqref{eq_power_scaling}, where critical parameter and exponents are also mentioned. In (b) and (c) we show the expected ergodic and Anderson transition point for \bte\ and RPE respectively.
	}
	\label{fig_3}
\end{figure}

In case of the RPE, bulk eigenvalues scale as: $\epsilon_\text{RPE} = \sqrt{4 + 2N/\sigma^2} = 2\sqrt{1 + N^{1 - \tilde{\gamma}}}$. Consequently the scaled DOS of RPE ($E\to E/\epsilon_\text{RPE}$) varies from Wigner's semicircle to Gaussian similar to \bte\ as $\tilde{\gamma}$ is increased from $0$. Contrarily for \xxx, DOS always follows a Gaussian distribution spreading with the disorder strength, which is typical of many-body systems with local interactions \cite{Brody2}. However, the differences in global shapes of DOS from these different models do not dictate the correlations in the respective local energy scales as demonstrated below.
\paragraph{\textbf{Nearest Neighbour Spacing (NNS):}} One of the most commonly investigated quantity reflecting local spectral correlations is the distribution of NNS of the ordered and unfolded eigenvalues (see Appx.~\ref{apnd_algo}), which follows Wigner's surmise and exponential distribution for the chaotic and integrable systems respectively \cite{Mehta1}. 
As the nature of correlation present in the spectrum edge is different than that of the bulk energy levels, we numerically evaluate the PDF of NNS choosing only middle 25\% of the spectrum. The results for $N = 8192$ are shown in Fig.~\ref{fig_2}(a) via markers along with the approximate empirical PDF of NNS (Eqs.~(12) and (13) in \cite{Caer1}), which is $N$-independent, provided $N\gg 1$. Such a functional form also implies that for $s\ll 1$, $\prob{\beta; s} \sim s^\beta\;\forall\; \beta$, thus the degree of level repulsion is indeed quantified via $\beta$, as expected from Eq.~\eqref{eq_beta_jpdf}. We observe a crossover from Wigner's surmise to exponential distribution in Fig.~\ref{fig_2}(a) as we decrease $\beta$, implying a suppression of chaos. A similar crossover is observed in RPE \cite{Pino1} and \xxx\ \cite{Buijsman1} as well.
\paragraph{\textbf{Ratio of Nearest Neighbour Spacing (RNNS):}}Another notable measure of the short-range spectral correlations is the RNNS, which is much simpler to study since unfolding of the energy spectrum is not required \cite{Atas1,Oganesyan1}.
If we define $\tilde{r_i} = \min\cbr{r_i, 1/r_i}$,  where $r_i$ is the $i^{th}$ RNNS, then $\prob{\tilde{r}} = 2\prob{r}\Theta\del{1-r}$, with $\Theta(x)$ being the Heaviside step function. We show PDF of $\tilde{r}$ via markers in Fig.~\ref{fig_2}(b) for $N = 8192$ along with the empirical PDF of RNNS (Eq.~(1) in \cite{Corps1}). Again a crossover w.r.t. $\beta$ is immediately apparent. For $N\gg 1$, the density of RNNS is system size independent, as shown in Fig.~\ref{fig_2}(c) for two values of $\beta$ while varying $N$. This can inferred from Fig.~\ref{fig_3}(a) as well, where the ensemble averaged values of $\tilde{r}$ as a function of $\beta$ collapse for different $N$ provided $N\gg 1$. In Figs.~\ref{fig_2}(d), (e) and (f), we show the density of $\tilde{r}$ for different values of $\gamma$ while varying $N$. 
For  any value of $\gamma < 0$ as $N\to\infty$ (i.e. $\beta\to \infty$), the energy levels are highly correlated and strongly repel each other. The increase in level repulsion with $N$ is shown in Fig.~\ref{fig_2}(f) for a fixed value of $\gamma = -0.3$. 
Exactly at $\gamma = 0$ (i.e. $\beta = 1$) the density of $\tilde{r}$ is independent of $N$ and matches that of GOE (Fig.~\ref{fig_2}(e)). 
On the other hand for any $\gamma>0$ and $N\to\infty$ (i.e. $\beta\to 0$), the energy levels become uncorrelated and clustered as in Poisson ensemble.
In Fig.~\ref{fig_2}(d) we show that the  density of $\tilde{r}$ converges towards Poisson expression as we increase $N$ for a fixed value of $\gamma = 0.3$.
The analyses above imply that the signatures of chaos in the short-range spectral correlations are lost as we lower the repulsion parameter $\beta$. Now we identify the exact nature of such a transition.
\paragraph{\textbf{Criticality in chaotic-integrable transition:}}For a fixed $\gamma$, the quick convergence of PDF of RNNS with system size (Fig.~\ref{fig_3}(a)) enables us to conclude that the $\mean{\tilde{r}}$ has one to one correspondence with $\beta$ when $N\gg 1$. $\beta$ increases with $N$ for any $\gamma<0$ (as $\beta = N^{-\gamma}$), hence $\mean{\tilde{r}}$ should also increase with $N$ and vice versa.
Figure~\ref{fig_3}(b) conforms to the above expectations suggesting a scaling hypothesis for $\mean{\tilde{r}}$. Let us assume that there exists a relevant correlation length $\Xi$ showing a power law divergence around a critical point, $\gamma_c$, i.e. $\Xi \sim (\gamma - \gamma_c)^{-\nu}$, where $\nu$ is a critical exponent. Then any quantity $A(\gamma, N)$ showing non-analytical behaviour close to $\gamma_c$ should behave as
\begin{align}
	\label{eq_power_scaling}
	A(\gamma, N)\propto f\del{(\gamma - \gamma_c) (\log N)^{1/\nu}}
\end{align}
where $f$ is a universal function and we assume $\Xi$ to scale with $\log N$ instead of $N$. Such a scaling ansatz valid for $2^{nd}$ order phase transition is shown to hold in case of the Kullback-Leibler divergence of RPE \cite{Pino1}. We collapse the crossover curves from different system sizes based on Eq.~\eqref{eq_power_scaling} (see Appx.~\ref{apnd_algo}) and obtain $\gamma_c = 0.0030$ and $\nu = 1.0316$ as shown in the inset of Fig.~\ref{fig_3}(b). Such a critical behaviour can also be inferred from the scale invariance of $\mean{\tilde{r}}$ w.r.t. $-\log\beta = \gamma\log N$ (Fig.~\ref{fig_3}(a)). Comparing this with Eq.~\eqref{eq_power_scaling}, we get $\gamma_c = 0$ and $\nu = 1$, which is consistent with our numerical analysis.

We show the $\mean{\tilde{r}}$ curves for different system sizes and chain lengths in Fig.~\ref{fig_3}(c) and (d) for RPE and Heisenberg model respectively. Again assuming a power law behaviour like Eq.~\eqref{eq_power_scaling}, we are able to collapse the data for RPE using $\tilde{\gamma}_c\approx 1.9750, \nu \approx 1.1359$.
For Heisenberg chain we assume that $A(\gamma, N)\propto f\del{(h - h_c) L^{1/\nu}}$ and get 
$h_c \approx 2.7696, \nu \approx 0.7842$. Note that the critical disorder strength found here corresponds to the middle 25\% of eigenspectrum, hence conforms to the energy density phase diagram of MBL transition present in the literature \cite{Luitz1}.

Thus we show that the chaotic-integrable transition in all three models is $2^{nd}$ order in nature. The crucial difference lies in the physical significance of these critical points. We observe that the chaotic-integrable transition occurs at $\gtat$ in the case of RPE, i.e. the energy states localize as soon as the energy levels start to cluster. Contrarily in case of \bte, chaotic-integrable transition occurs at $\gamma = 0$ which we previously argued to be $\get$, i.e. where ergodicity breaks down. Thus in the thermodynamic limit ($N\to\infty$), there will be extended states for which energy levels are uncorrelated, which has a profound implication in the dynamical properties of \bte\ (Sec.~\ref{sec_dynamics}). Thus our analysis shows that the eigenstate localization property is not necessarily indicative of the degree of repulsion present in the energy spectrum as also observed in certain structured matrix ensembles \cite{Tang1, Mondal1, Mondal2}.
\begin{figure}[t]
	\centering
	\includegraphics[width=0.7\textwidth, trim = {0 190 0 195}]{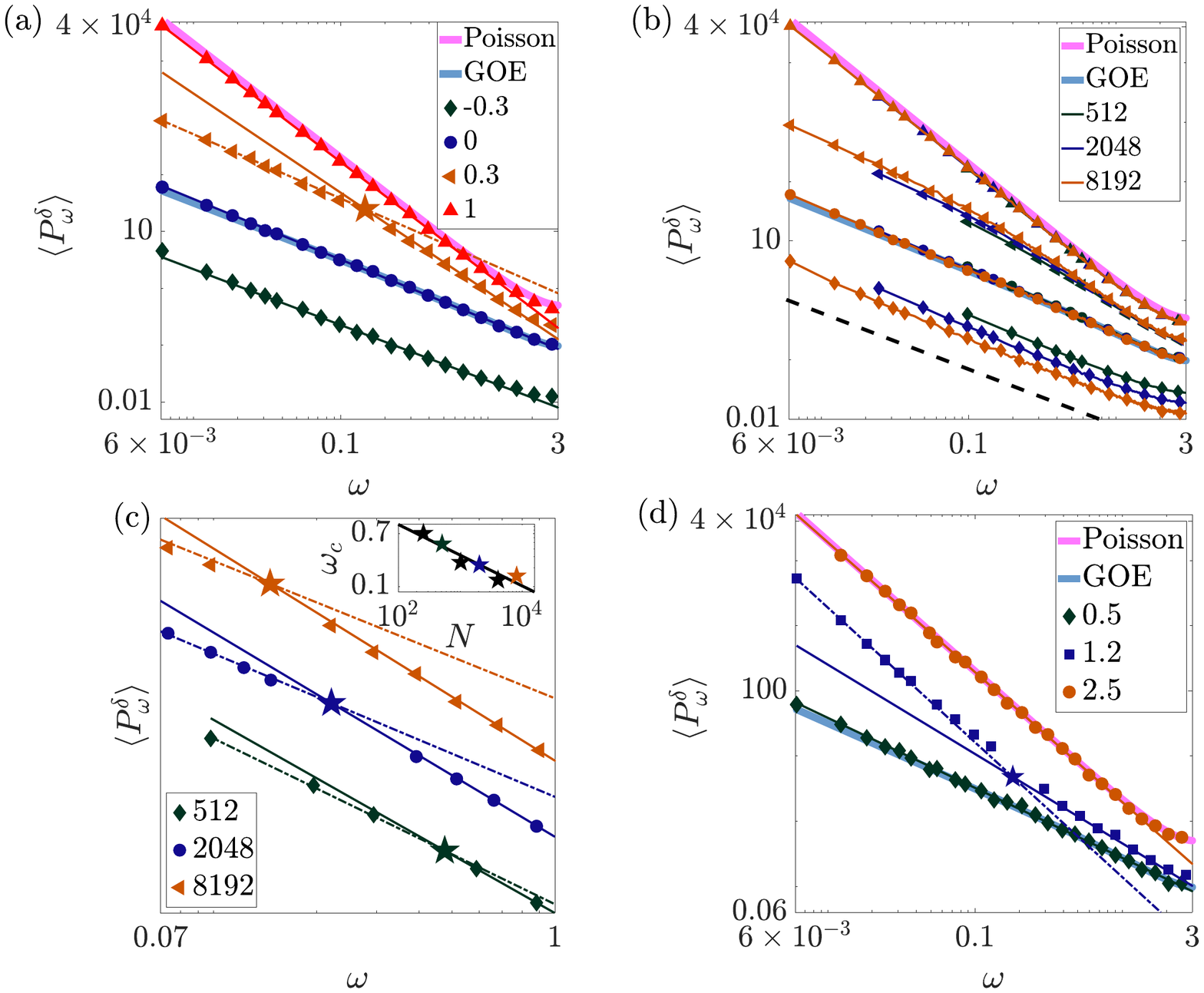}
	\caption{\textbf{Power spectrum of noise} for (a) \bte\ and (d) RPE for $N = 8192$ as a function of dimensionless frequency, $\omega = 2\pi k/N$. In (b), power spectrum is shown for various $\gamma$ (denoted by markers similar to (a)) and $N$ (denoted by different colors as in the legend). We see that $\mean{P_\omega^\delta}$ shifts downwards with increasing $N$ for $\gamma = -0.3$, where the dashed line $\propto 1/\omega$ is placed as a guide to the eye.
		(b) $\mean{P_\omega^\delta}$ for $\gamma = 0.3$ while varying $N$ (data shifted in Y-direction for clarity). Dashed and solid lines indicate $1/\omega$ and $1/\omega^2$ behaviors respectively. In (a), (c) and (d), stars denote the critical frequencies, $\omega_c$ separating heterogeneous behavior. Inset of (c) shows numerically obtained $\omega_c$, where solid line denotes linear fit in log-log scale. We also show the analytical $\mean{P^\delta_\omega}$ for Poisson and GOE (Eq.~(10) in \cite{Faleiro1}) via bold curves.
	}
	\label{fig_4}
\end{figure}
\paragraph{\textbf{Power spectrum:}}Short-range spectral correlations in \bte\ exhibit criticality only around $\get = 0$. We also expect a second critical point associated with the localization transition, which can be captured by the long-range spectral correlations, e.g. power spectrum of $\delta_n$ statistics defined as \cite{Faleiro1, Riser1},
\begin{align}
	P_k^\delta = |\hat{\delta_k}|^2,\quad \hat{\delta_k} = \frac{1}{\sqrt{N}}\sum_{n}\delta_n\exp\del{-\frac{i2\pi kn}{N}} 
\end{align}
where $\delta_n\equiv \mathcal{E}_n - n$ is the fluctuation of the $n^{th}$ unfolded energy level, $\mathcal{E}_n$ around its mean value, $n$. Ensemble average of $P_k^\delta$, denoted by $\mean{P_k^\delta}$, is explored for \xxx\ in \cite{Corps2, Corps3}. For $\gamma\geq 0$, there exists a critical frequency $k_c = N^{1-\gamma}/2$ in \bte\ \cite{Relano3} such that for $k\leq k_c, \mean{P_k^\delta}\propto 1/k$ identifying completely chaotic behaviour whereas for $k>k_c, \mean{P_k^\delta}\propto 1/k^2$, which is a signature of Poisson ensemble.
Note that the power spectrum of some physical systems like Robnik billiard \cite{Gomez2}, kicked top \cite{Santhanam2} exhibit a homogeneous behaviour, $\mean{P_k^\delta}\propto 1/k^\alpha$, across all frequencies with $1<\alpha<2$.

In Fig.~\ref{fig_4}(a), we show the power spectrum of \bte\ as a function of dimensionless frequency, $\omega = 2\pi k/N$ for $N = 8192$ and various $\gamma$ with the bold curves showing the analytical forms of $\mean{P^\delta_\omega}$ for Poisson and GOE (Eq.~(10) in \cite{Faleiro1}). In Fig.~\ref{fig_4}(b), we show $\mean{P_\omega^\delta}$ for same values of $\gamma$ but also by varying $N$ denoted by different colors. We see that for finite $N$ and $\gamma<0$, $\mean{P_\omega^\delta}\propto 1/\omega$ for a typical value of $\gamma = -0.3$. 
However, we observe that for $N\gg1$ and $\gamma\ll 0$ (i.e. $\beta\to \infty$), there are fluctuations around $1/\omega$ behavior due to the energy spectrum attaining a picket-fence structure. For $\gamma\geq 0$, we can identify two critical points separating three distinct regimes by looking at Fig.~\ref{fig_4}(b) or from the analytical calculations in \cite{Relano3}:
\begin{itemize}
	\item $\gamma = 0$: $\mean{P_\omega^\delta}\propto 1/\omega$ for any $N \Rightarrow$ energy levels are correlated at all scales even in the thermodynamic limit
	\item $0 <\gamma < 1$: Heterogeneous spectra  $\Rightarrow \omega_c = 2\pi k_c/N = \pi N^{-\gamma}$ separating Poisson and GOE like scaling. In Fig.~\ref{fig_4}(c), we show $\mean{P_\omega^\delta}$ for $\gamma = 0.3$ and various $N$, which clearly reflects the heterogeneous features. In the inset we show numerically obtained $\omega_c$ vs. $N$.
	\item $\gamma\geq 1$: $ \mean{P_\omega^\delta}\propto 1/\omega^2$ for any $N$ $\Rightarrow$ energy levels are uncorrelated at all scales even in the thermodynamic limit
\end{itemize}
Note that $k_c\to \infty$ for $N\to \infty$ and $0<\gamma<1$, i.e. signature of chaotic spectrum prevails over infinitely many frequencies. However, their support set constitute a zero fraction of the set of principle frequencies as $k_c/k_\text{Nyquist} = N^{-\gamma}\to 0$ for any $\gamma>0$ ($k_\text{Nyquist} \approx N/2$ is the highest frequency required to fully reconstruct the original spectrum \cite{Faleiro1}). Such a fractal behaviour suggests the absence of ergodicity in the \bte\ for $0<\gamma<1$. For example, in case of RPE, eigenstates occupy zero fraction of the Hilbert space volume despite being extended in the NEE phase ($1\leq \tilde{\gamma}<2$) \cite{Kravtsov1}. Corresponding power spectrum also exhibits heterogeneous behaviour as shown in Fig.~\ref{fig_4}(d).
Thus we can attribute the heterogeneity in power spectrum of $\delta_n$ statistics to the existence of NEE phase and we can conclude that \bte\ enters the NEE phase for $0<\gamma<1$ where ergodicity breaks down at $\get = 0$ and Anderson transition occurs at $\gat = 1$.

With this primary evidence of the existence of NEE regime in \bte, in the next section we will study the eigenfunction properties and obtain the fractal scaling of NEE states.

\section{Properties of Eigenstates}\label{sec_state}
Due to the canonical invariance, the eigenvectors of $N\times N$ GOE matrices are uniformly distributed in the unit $N$-dimensional sphere, resulting in mutually independent eigenvector components. Contrarily for \bte, all elements but the first component of the $k^{th}$ eigenvector can be expressed in terms of the $k^{th}$ eigenvalue and different matrix elements \cite{Dumitriu1}. Hence even for typical values of $\beta$ (i.e. $\beta = 1, 2, 4$), the eigenvector properties of Wigner-Dyson and \bte\ are different from each other, although their energy level statistics are identical. This can be readily verified from the distribution of $\log(N\abs{\Psi_i}^2)$ ($\Psi_i$ is $i^{th}$ component of the eigenstate $\ket{\Psi}$), which has a long tail for $\beta = 1$ in \bte\ compared to GOE.
\begin{figure}[t]
	\centering
	\includegraphics[width=\textwidth, trim = {0 175 0 180}]{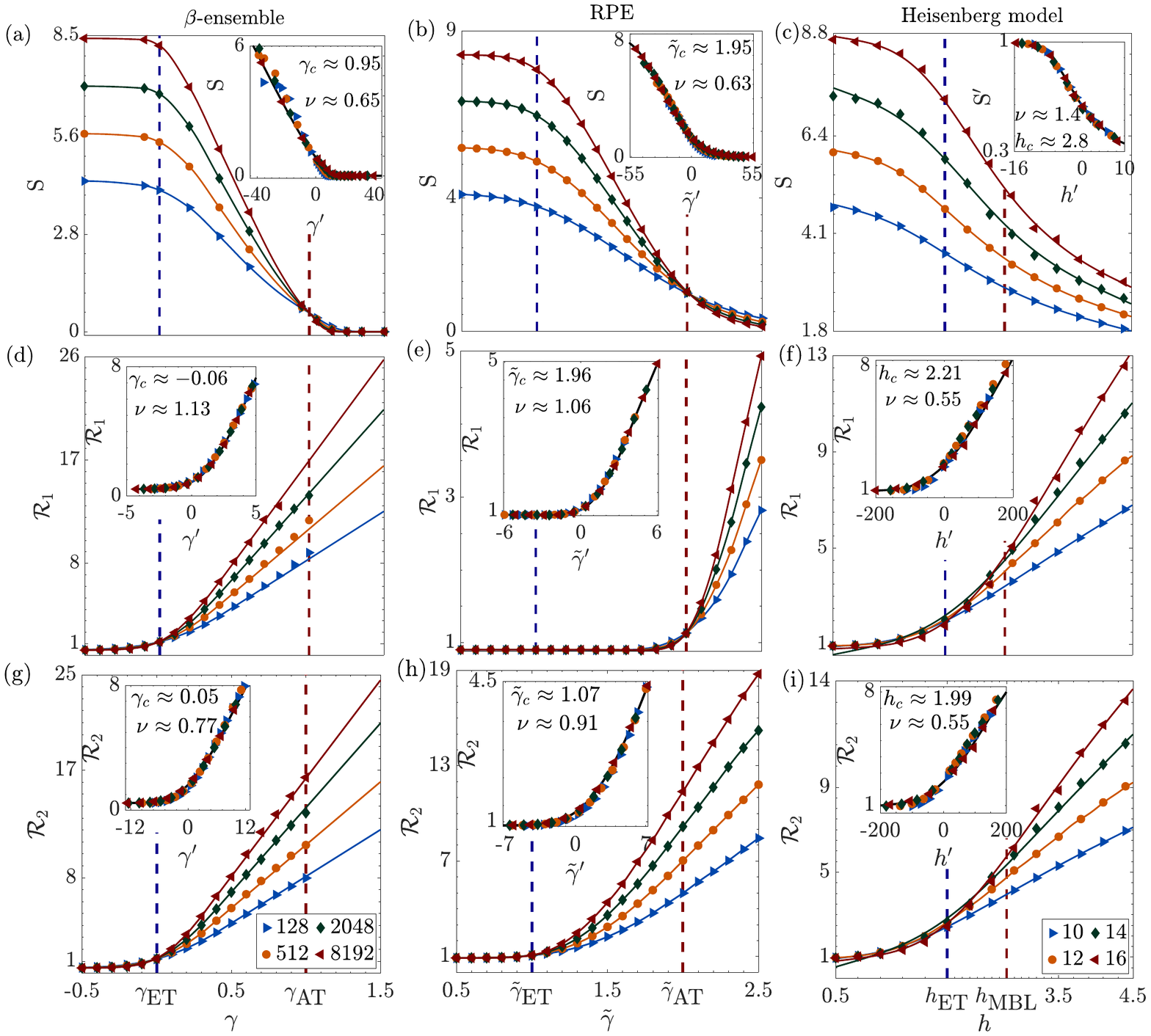}
	\caption{\textbf{Eigenstate statistics:} Shannon entropy, $\shn$ and relative R{\'e}nyi entropy of two types ($\mathcal{R}_{1, 2}$ in Eq.~\eqref{eq_RR}) for \bte, RPE and \xxx, as a function of system parameters for different matrix size, $N$ and chain length, $L$ (values given in the legends). The critical values of parameters indicating ergodic and localization transitions are marked by dashed line in all plots. Inset shows collapsed data following the ansatz in Eq.~\eqref{eq_power_scaling}, where numerically obtained critical parameters and exponents are also given. Inset of (c) shows collapsed data of $\shn' = \shn/\log(0.48 N)$.
	}
	\label{fig_5}
\end{figure}
\paragraph{\textbf{Localization transition:}} In order to characterize the Anderson transition from the properties of eigenstates, we begin by computing the Shannon entropy, defined as, $\shn = -\sum_{i = 1}^{N}P_i\log\del{P_i}$ with $P_i = |\Psi_i|^2$.
In Fig.~\ref{fig_5}(a), we show ensemble averaged $\shn$, obtained from the eigenstates taken from middle 25\% of the spectra, exhibiting a non-analyticity around $\gamma = 1$. Assuming a power-law behaviour of the relevant correlation length, we obtain critical parameter $\gamma_c = 0.95$ and exponent $\nu=0.65$ using Eq.~\eqref{eq_power_scaling}, while the collapsed data is shown in the inset of Fig.~\ref{fig_5}(a). We also observe that the Inverse Participation Ratio (IPR), I $= \sum_{i = 1}^{N}\abs{\Psi_i}^4$ exhibits a criticality around $\gamma = 1$ as shown in Fig.~\ref{fig_6}. Thus we confirm that the Anderson transition occurs at $\gat \equiv 1$ for \bte, while at $\gtat \simeq 2$ for RPE [Fig.~\ref{fig_5}(b)].

For \xxx, Shannon entropy is almost constant for a particular $L$ if disorder strength is small ($h\ll 1$) and slowly decaying for $h\gg 1$. Similar behaviour for IPR indicates that the energy states of the Heisenberg model in the MBL regime are extended in the Hilbert space exhibiting a non-trivial multifractal behaviour \cite{Luitz1}. 
However according to \cite{Torres2}, one may look at the ratio of Shannon entropies of Heisenberg model and GOE, i.e. $\shn' = \shn/\shn_\text{GOE}\approx \shn/\log(0.48 N)$, where $N = {L\choose L/2}$ is the Hilbert space dimension of the $S_z = 0$ symmetry sector. 
The finite-size scaling of $\shn'$ gives the numerical estimate of the MBL transition point to be $h_\text{MBL}\approx 2.77$ for our choice of $L=8,10,\cdots,16$.
This is the same critical point beyond which energy levels start to cluster [Fig.~\ref{fig_3}(d)]. Such a conclusion is also verified via studies of entanglement entropy, magnetization fluctuations \cite{Luitz1}. Thus unlike \bte, eigenstates start to localize as soon as energy levels begin to cluster for both RPE and \xxx.
\begin{figure}[t]
	\centering
	\includegraphics[width=0.32\textwidth, trim = {0 25 0 25}]{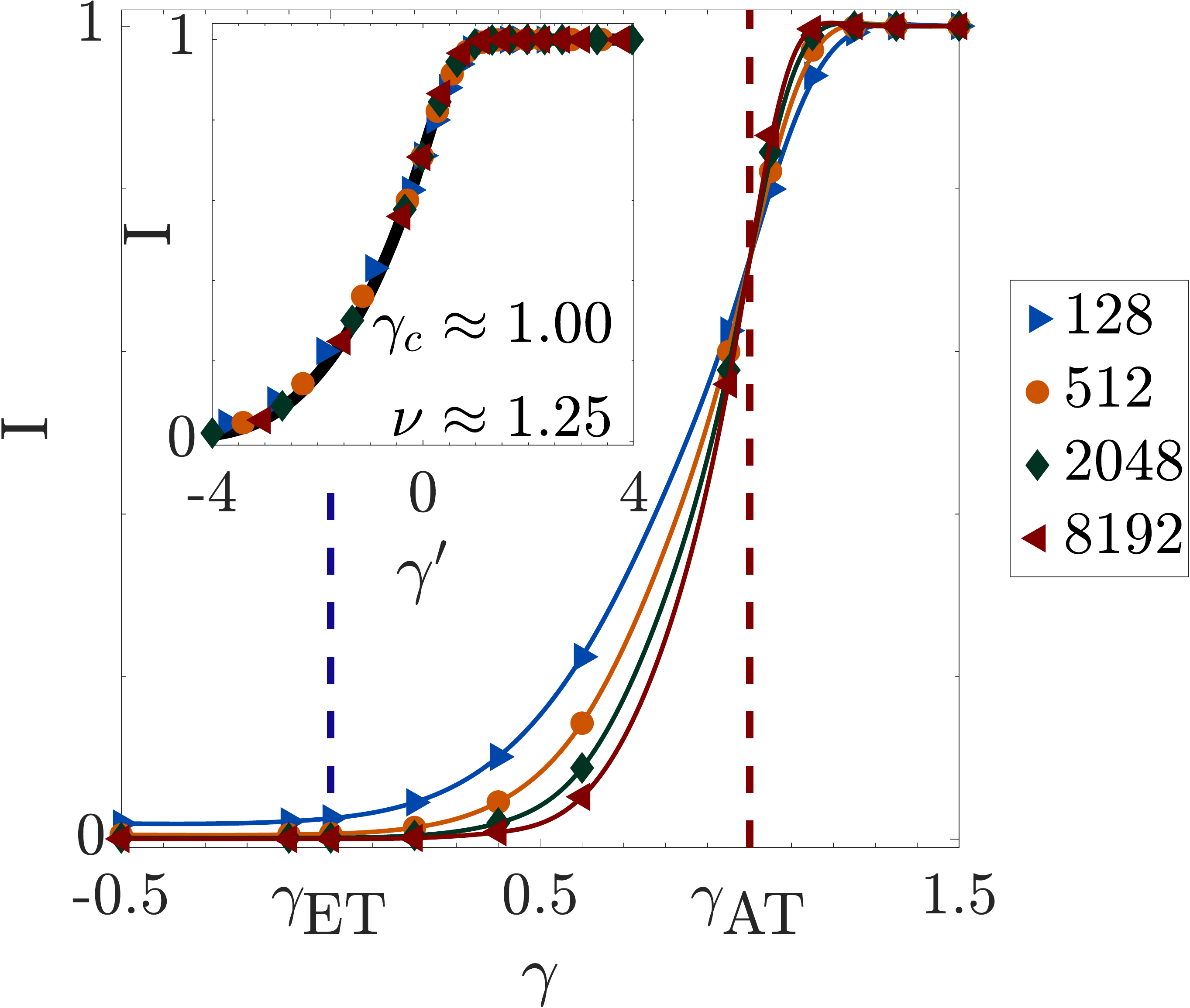}
	\caption{\textbf{Inverse Participation Ratio:} for \bte\ as a function of $\gamma$ for various system sizes, $N$. Inset shows collapsed data following the ansatz in Eq.~\eqref{eq_power_scaling} along with critical parameter and exponent.
	}
	\label{fig_6}
\end{figure}
\paragraph{\textbf{Ergodic to non-ergodic transition:}}
We now quantify the loss of ergodicity by computing the relative R{\'e}nyi ($\mathcal{R}$) entropy between a pair of eigenfunctions \cite{Nagy2} having similar energy densities. Let $\ket{\Psi_i^j}$ be the $i^{th}$ eigenvector of the $j^{th}$ disordered realization of an ensemble. We define two kinds of $\mathcal{R}$ as follows:
\begin{align}
	\label{eq_RR}
	\begin{split}
		\ra &= -2\log\del{\sum_{k = 1}^N \abs{\Psi_i^j(k) \Psi_{i+1}^j(k)}},\qquad
		\rb = -2\log\del{\sum_{k = 1}^N \abs{\Psi_i^j(k)\Psi_{i+1}^{j'}(k)}}.
	\end{split}
\end{align}
Here $\ra$ and $\rb$ measure similarity among wavefunctions obtained from the same and different samples respectively. For eigenstates of GOE, $\mathcal{R} = -2\log\del{\sum_{k = 1}^N |z_k|},\: z_k = x_k\times y_k$, where $x_k, y_k$ are i.i.d. random variables and for $N\gg 1$, $x_k, y_k\sim \mathcal{N}(0, 1/\sqrt{N})\;\forall\; k$ assuming wavefunctions are normalized. Then, $\prob{z} = N\fbslk{0}{N\abs{z}}/\pi$, where $\fbslk{0}{x}$ is the modified Bessel function of $2^{nd}$ kind. Then,
\begin{align}
	\label{eq_R_goe}
	\begin{split}
		\mean{\abs{z}} = \dfrac{2}{N\pi}\Rightarrow \mean{\mathcal{R}} = -2\log\del{\sum_{i = 1}^N \dfrac{2}{N\pi}} = 2\log\del{\dfrac{\pi}{2}} \approx 0.9032
	\end{split}
\end{align}
Thus $\mathcal{R} \approx 0.9$ for any pair of completely extended wavefunctions and this value is used to benchmark our numerical estimates. The relative R{\'e}nyi entropy can be viewed as a generalization of the Kullback-Leibler divergence exhibiting critical behaviour in the case of RPE \cite{Pino1}. We will investigate $\ra$ and $\rb$ in the similar spirit with the premises of finding:
\begin{itemize}
	\item Ergodic regime: $\mathcal{R}$ $\sim \mathcal{O}(1)$ for any pair of wavefunctions, as both of them are uniformly extended 
	\item Localized regime: $\mathcal{R}$ will diverge as different wavefunctions localize at separate sites
	\item Non-ergodic regime having two possibilities:
	\begin{itemize}
		\item If energy levels of $\ket{\Psi_i^j}, \ket{\Psi_{i'}^j}$ repel each other, then such energy states come from the same symmetry sector, i.e. the same subspace of the Hilbert space. Thus $\ket{\Psi_i^j}, \ket{\Psi_{i'}^j}$ are likely to hybridize if the governing Hamiltonian is sufficiently dense \cite{Pino1}, giving $\mathcal{R}$ $\sim \mathcal{O}(1)$.
		\item In absence of any level repulsion, the energy states in the NEE phase are likely to be extended over different parts of the Hilbert space, thus $\mathcal{R}$ will diverge.
	\end{itemize}
\end{itemize}
Let us now illustrate the measures for the well studied case of RPE. For two nearby energy states $\ket{\Psi_i^j}, \ket{\Psi_{i'}^j}$ with comparable energy densities, $\ra \sim \mathcal{O}(1)$ for $\tilde{\gamma}<\gtat$ and $\ra \gg 1$ for $\tilde{\gamma}>\gtat$ as chaotic-integrable transition occurs at $\gtat$.
On the other hand, the energy states from different samples, say $\ket{\Psi_i^j}, \ket{\Psi_{i'}^{j'}}$, are likely to have different support set in the NEE phase, as different governing Hamiltonians cannot hybridize them even if their energy densities are comparable giving $\rb\gg 1$ for $\tilde{\gamma}>\tilde{\gamma}_\text{ET}$. In Fig.~\ref{fig_5}(e) and (h) we show $\ra, \rb$ for RPE exhibiting $2^{nd}$ order phase transitions clearly identifying the critical points $\tilde{\gamma}_\text{AT}$ and $\tilde{\gamma}_\text{ET}$ respectively.

Recall that for \bte\ chaotic-integrable transition occurs at $\get = 0$, hence $\ra$ should show non-analyticity at the same point. Previously we argued that $\get$ is also the ergodic transition point, thus $\rb$ should exhibit criticality there as well. The critical behaviours of $\ra, \rb$ are evident in Fig.~\ref{fig_5}(d) and (g), where scaling analysis gives $\gamma_c \equiv \get \sim 0$.

In the case of \xxx, we find that the scaling of $\rb$ indicate a $h_c\equiv h_\text{ET}\approx 2$ in Fig.~\ref{fig_5}(i) while earlier we identified $h_\text{MBL}\approx 2.77$. This indicates the existence of NEE in \xxx\ for an intermediate range of disorder $h\in (2, 2.77)$ in agreement with the existing study on participation entropy, survival probability \cite{Torres2} and momentum distribution fluctuations \cite{Corps2}. Therefore one would expect $\ra$ to show criticality at $h_\text{MBL}$ since it is also the chaotic-integrable transition point. However the Hamiltonians of \xxx\ are so sparse that they fail to completely hybridize the NEE eigenstates even from the same subspace of the Hilbert space. As a result $\ra$ shows criticality at $h_c\approx 2.21$ [Fig.~\ref{fig_5}(f)], a value in between $h_\text{ET}$ and $h_\text{MBL}$. Below the critical values, $\mathcal{R}\approx 0.9$ in all three models as expected from Eq.~\eqref{eq_R_goe}.
\begin{figure}[t]
	\centering
	\includegraphics[width=\textwidth, trim = {0 355 0 355}]{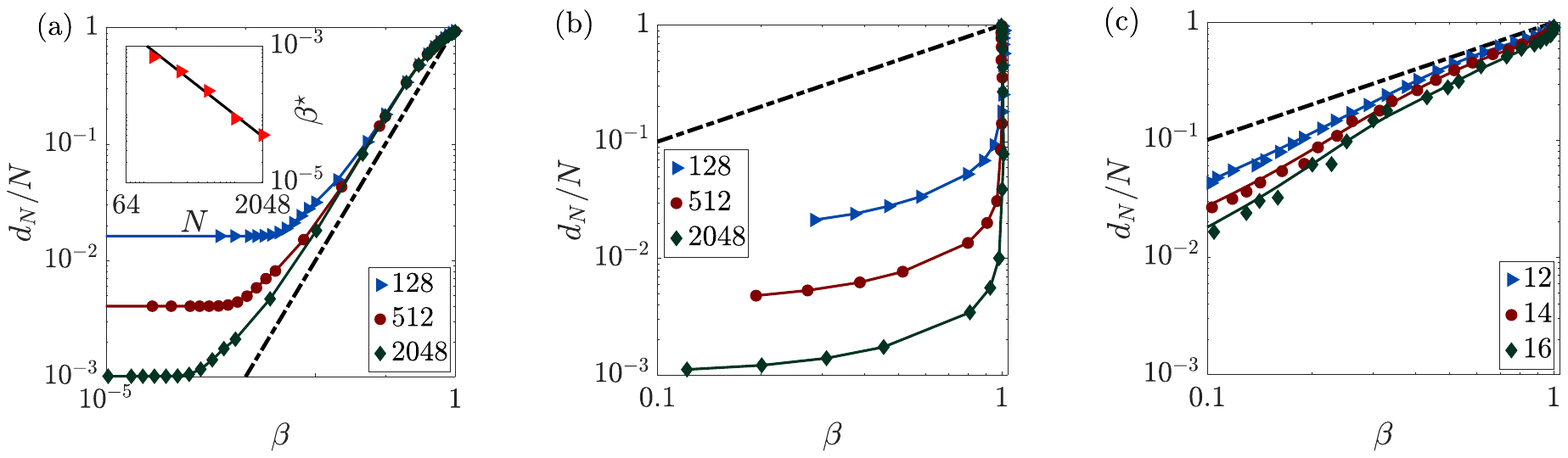}
	\caption{\textbf{Entropic localization length} for \bte, RPE and \xxx\ for various system sizes and chain lengths as a function of repulsion parameter $\beta$. We show the line $d_N/N = \beta$ via dashed curve. Inset of (a) shows $\beta^\star$ below which $d_N/N$ becomes constant.
	}
	\label{fig_7}
\end{figure}
\begin{figure}[b]
	\centering
	\includegraphics[width=\textwidth, trim = {0 345 0 350}]{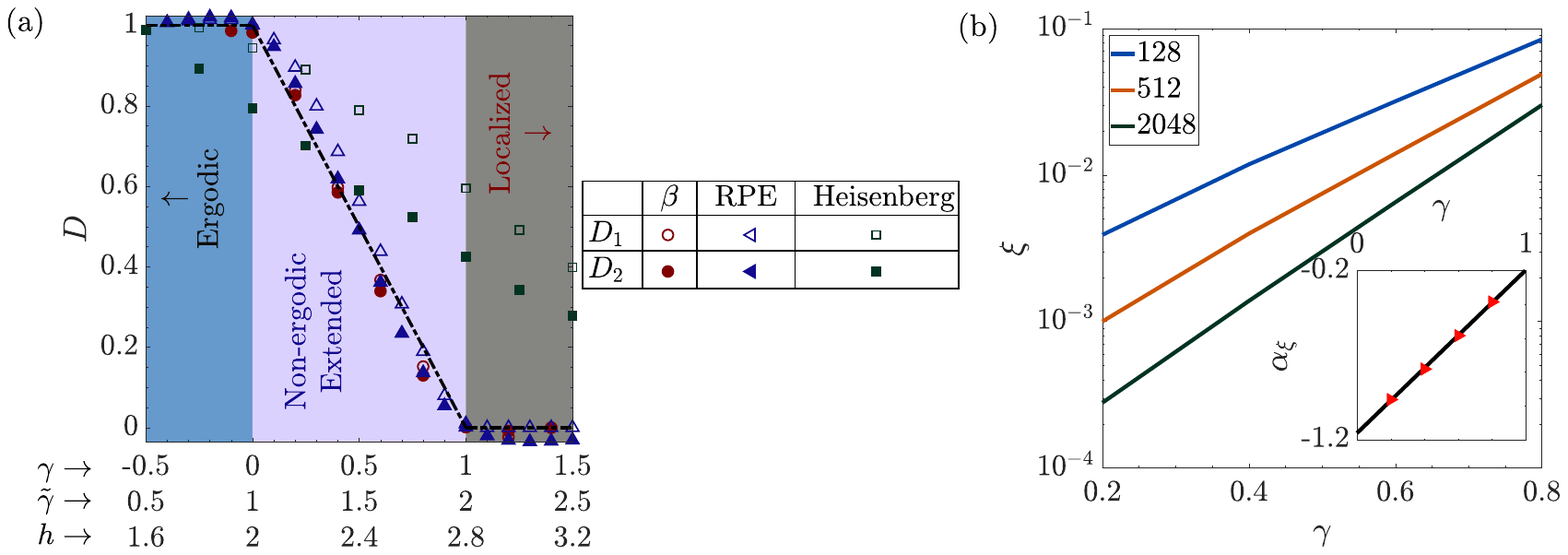}
	\caption{(a) \textbf{Phase diagram:} The three distinct phases observed in \bte, RPE and \xxx\ are demarcated against the critical parameter values. Markers indicate fractal exponents $D_{1, 2}$ as explained in the legend. (b) \textbf{Fraction of localized states:} as a function of $\gamma$ for \bte\ for various system sizes. Inset shows $\alpha_\xi$, the system size scaling exponent of $\xi$ vs. $\gamma$ along with a linear fit, $\alpha_\xi = a\gamma + b$, where $a = 0.9656\pm 0.0676$ and $b = -1.1618\pm 0.0370$.
	}
	\label{fig_8}
\end{figure}
\paragraph{\textbf{Localization length:}}
In the previous section, we noticed that the spectral properties of \bte\ and RPE have an important difference: chaotic-integrable transition occurs at $\get$ for \bte\ and $\gtat$ for RPE. The degree of level repulsion, $\beta$ has been interpreted as the rescaled localization length in various systems \cite{Sorathia1,Izrailev3,Casati2,Flores1} though there are exceptions as well \cite{Benito-Matias1}. Now we look at the entropic localization length w.r.t. the Shannon entropy, $d_N \equiv 2.07e^\shn$, such that $d_N\approx N$ or $d_N\approx 1$ for a fully ergodic or localized energy state respectively \cite{Sorathia1}.
In case of RPE or \xxx, one can numerically fit the PDF of NNS with any phenomenological model (e.g. Brody \cite{Brody1}, Berry-Robnik \cite{Berry2} etc.) to estimate the repulsion parameter, $\beta$. However, such a numerical fit pertains to the global shape of the PDF of NNS \cite{Sorathia1}, without necessarily reflecting the behaviour of $\prob{s}_{s\to 0}$, which is the true measure of level repulsion in a system. Thus we exploit the one to one correspondence between $\beta$ and the mean value of RNNS. We observe a sub-linear behaviour when $\beta$ is small and $d_N/N$ converges to 1 when $\beta\to 1$ [Fig.~\ref{fig_7}(b) and (c)]. Hence energy states are completely extended whenever the energy levels repel each other and this justifies the coincidence of chaotic-integrable transition with the delocalization-localization transition in RPE and \xxx.

\begin{figure}[t]
	\centering
	\includegraphics[width=\textwidth, trim = {0 335 0 335}]{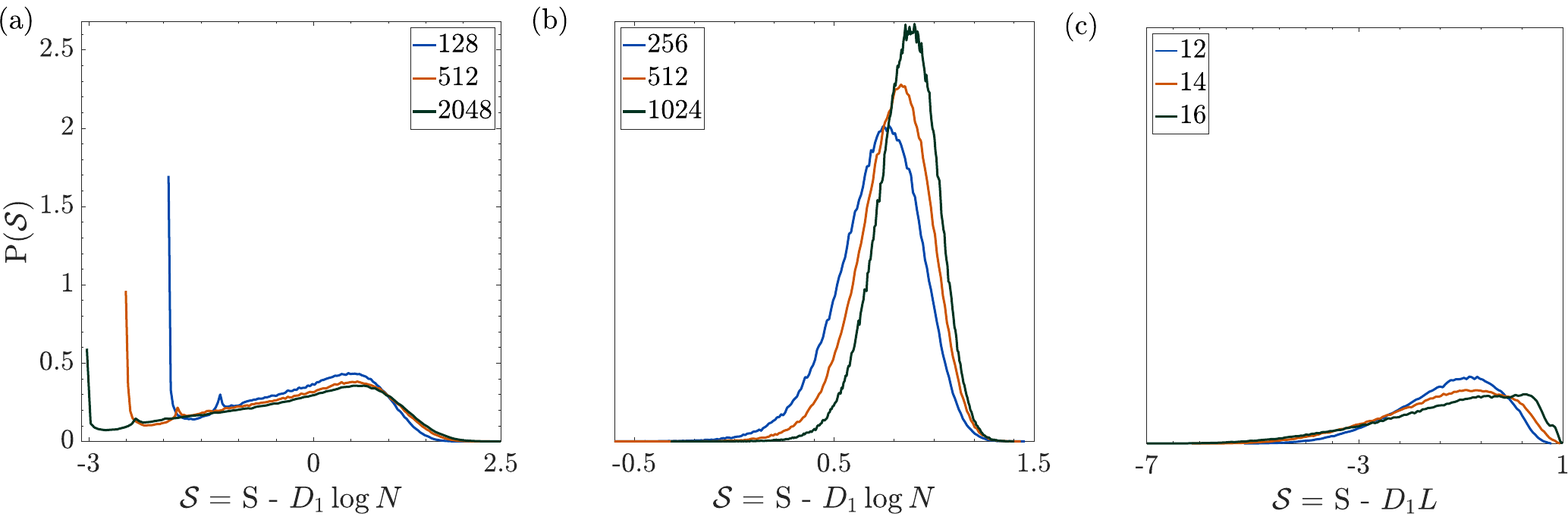}
	\caption{\textbf{Density of Shannon entropy:} for (a) \bte, $\gamma = 0.6$ (b) RPE, $\tilde{\gamma} = 1.5$ and (c) Heisenberg model, $h = 2.4$ while varying system sizes, $N$ and chain lengths, $L$, where $\mathcal{S}$ and S are the scaled and original Shannon entropies respectively ($D_1$ is the fractal dimension obtained from system size scaling of S.)
	}
	\label{fig_9}
\end{figure}

We show $d_N/N$ as a function of $\beta$ for various system sizes in case of \bte\ in Fig.~\ref{fig_7}(a). Here the relationship between the localization length and the degree of level repulsion is super-linear throughout. Moreover, $d_N/N$ becomes independent of $\beta$ for $\beta\leq \beta^\star\ll 1$ while $\beta^\star\propto 1/N$ as shown in the inset of Fig.~\ref{fig_7}(a). This implies that the localization transition should occur roughly at $\beta = 1/N$ as chaotic-integrable transition occurs at $\beta = N^{\get} = 1$. This supports our earlier observation that $\gat = 1$ does not coincide with the chaotic-integrable transition point in the case of \bte\ unlike RPE or Heisenberg model.
\paragraph{\textbf{Scaling of eigenstate fluctuations:}}It is important to analyse the eigenfunction fluctuations which can be quantified via R{\'e}nyi entropy, $S_R(q, N) \sim N^{D_q}$ where $D_q$'s are {\it fractal} dimensions for different values of $q$ \cite{Atas3}. For $q = 1$, the R{\'e}nyi entropy converges to the Shannon entropy, $\shn\sim D_1\log N$. Similarly for $q = 2$, one obtains the scaling in IPR $\sim N^{-D_2}$. In the ergodic regime the fractal dimensions, $D_{1, 2} = 1$ as eigenstates occupy the full Hilbert space volume, while $D_{1, 2} = 0$ in the localized regime. In the NEE phase, $0<D_{1, 2}<1$, which implies that the eigenstates are extended over infinitely many but a zero fraction of all possible sites in the thermodynamic limit (i.e. $N^{D_{1, 2}}\to \infty$ but $N^{D_{1, 2}}/N\to 0$ if $N\to\infty$). Since distributions of Shannon entropy and IPR are quite broad and skewed (Fig.~\ref{fig_9}(a)), median instead of mean has been used to estimate such fractal dimensions \cite{Mirlin1}.
The numerically estimated $D_{1, 2}$ clearly identifies the ergodic, NEE and localized regimes for \bte\ and RPE as shown in Fig.~\ref{fig_8} ($D_{1, 2}\approx 1-\gamma$ and $2-\tilde{\gamma}$ in the NEE phase for \bte\ and RPE respectively). However, in the case of Heisenberg model, $D_{1, 2}$ does not vanish for $h > h_\text{MBL}$ resulting from non-trivial multifractality in the MBL phase \cite{Luitz1}.

To probe finer details of the eigenstructure of \bte, we look at the density of Shannon entropy. The distributions from different system sizes collapse on top of each other at $\gamma = 0, 1$ upon a rescaling $\shn\to \shn - D_1\log N$. However, in the NEE phase ($0<\gamma<1$), two peaks emerge in the histogram of $\shn$: i) a broad peak whose location roughly scales as $N^{-D_1}$, ii) a sharp peak at $\shn = 0$, whose height decreases with $N$ (Fig.~\ref{fig_9}(a)). The existence of the $2^{nd}$ peak (which is absent for RPE, Fig.~\ref{fig_9}(b) and Heisenberg model, Fig.~\ref{fig_9}(c)) implies that a small albeit finite fraction of eigenstates are localized for $0<\gamma<1$. In Fig.~\ref{fig_8}(b), we show $\xi$, the fraction of localized eigenstates as a function of $\gamma$ for various $N$. Inset of the same figure shows $\alpha_\xi\approx \gamma - 1$, the system size scaling exponent of $\xi$, which implies roughly $\xi\propto N^{\gamma - 1}$. 
Thus in the thermodynamic limit, there will be an infinite number of completely localized states in the intermediate regime of \bte\ (since $N^\gamma\to \infty$ for $\gamma>0$), which constitutes a zero fraction of all possible eigenstates (since $\xi\to 0$ for $\gamma<1$). By looking at the median and mode of IPR of individual eigenstates for different system sizes, we find that the high energy states (i.e. the ones in the middle of spectrum) have a greater tendency to be localized compared to the ones close to the ground state in the NEE regime. Thus contrary to RPE, \bte\ offers two kinds of eigenstates in the NEE phase: $N^\gamma$ number of completely localized and $(N - N^\gamma)$ number of NEE states.
\section{Properties of Dynamics}\label{sec_dynamics}
So far we have looked at the statistical picture of the energy level correlation, eigenstate localization end ergodic properties of \bte. Now we want to look at the dynamical aspects of \bte, revealing important time and energy scales.
In this regard, one of the largest time scales is the Heisenberg time, defined as the inverse of mean level spacing. 
Beyond such a time, the energy level dynamics of a system equilibrates e.g. the spectral form factor attains a stationary state \cite{Suntajs1}.
Now we explicitly look at the time evolution of an initially localized state and identify the relevant dynamical timescales.
\paragraph{\textbf{Survival probability:}}
An important characterization of the dynamics of a quantum mechanical system is often done by monitoring the time evolution of a given wavefunction. We choose a unit vector $\ket{j}$ having energy close to the spectrum centre of $H$ as our initial state. 
Let $\del{E_k, \ket{\phi_k}}$ be the $k^{th}$ eigenpair of $H$ such that the time evolution of the initial state is given by,
\begin{align}
	\label{eq_time}
	\ket{j(t)} = e^{-iHt}\ket{j} = \sum_{k} e^{-iE_kt}\phi_k^{(j)}\ket{\phi_k},\quad \phi_k^{(j)} = \braket{\phi_k|j}.
\end{align}
The spread of initial state $\ket{j}$ over all other states is controlled by the off-diagonal terms in $H$ and is quantified by
the survival probability \cite{Schiulaz1}
\begin{align}
	\label{eq_survival}
	R(t) &= \abs{\braket{j|j(t)}}^2 = \abs{\sum_{k=1}^N \abs{\phi_k^{(j)}}^2 e^{-iE_kt}}^2
\end{align}
\begin{figure}[t]
	\vspace{-10pt}
	\includegraphics[width=\textwidth, trim=0 260 0 260]{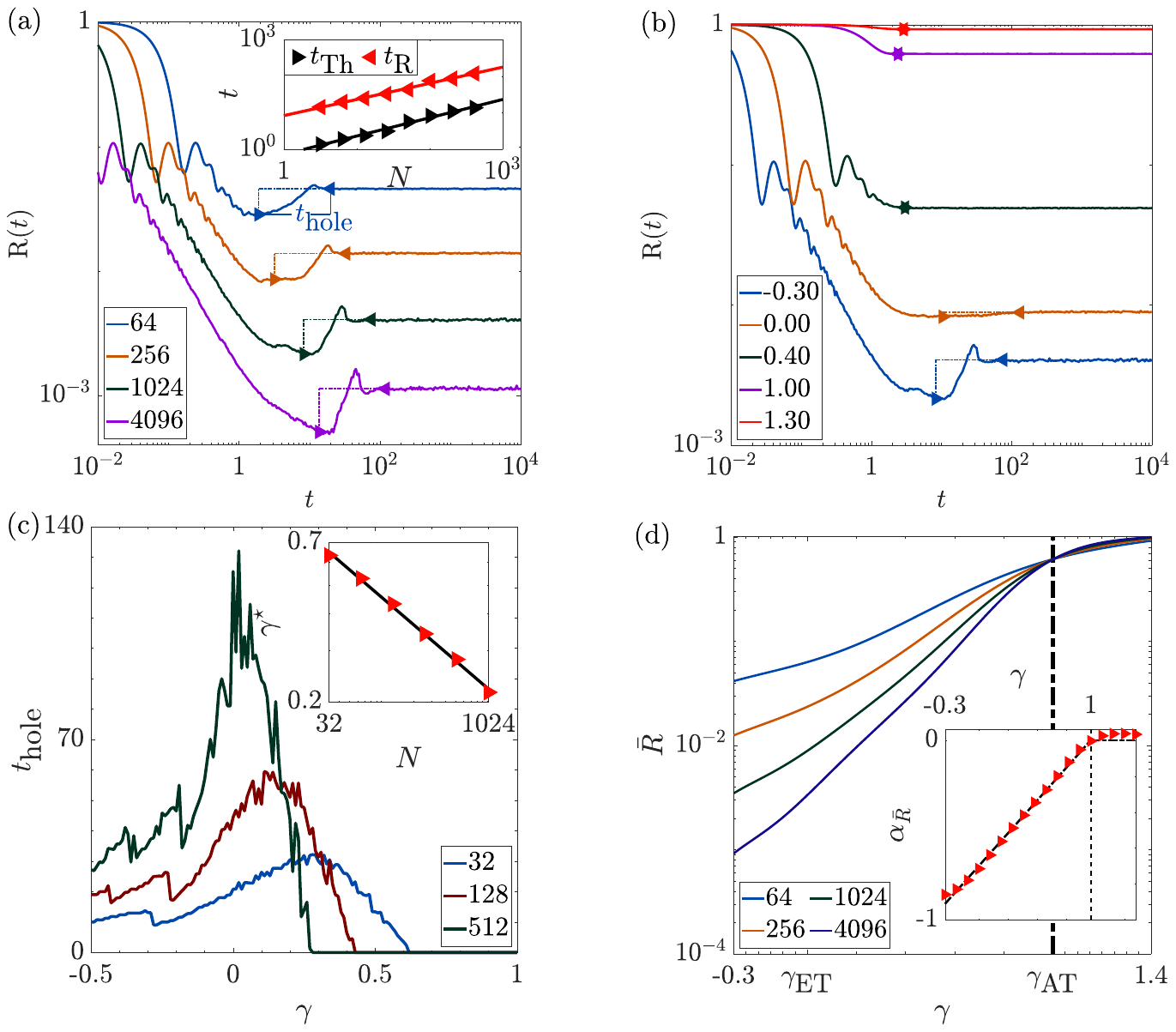}
	\caption[]{\textbf{Survival Probability for \bte:} (a) time evolution for $\gamma = -0.3$ and various system size, $N$. We show Thouless ($\tTh$), relaxation time ($\tR$) via markers and correlation hole ($t_\text{hole}$) via dashed line in each case. Inset shows $\tTh$ and $\tR$ as a function of $N$ along with linear fit in log-log scale via solid line, indicating a $\sqrt{N}$ dependence.
		(b) $N = 1024$ varying $\gamma$ 
		(c) Correlation hole vs. $\gamma$ for various $N$. Solid lines show smooth trend of the data using spline interpolation. Inset shows $\gamma^\star$ vs. $N$ in log-log scale ($t_\text{hole}\approx 0$ for $\gamma\geq \gamma^\star$).
		(d) Asymptotic value of survival probability vs. $\gamma$ for different $N$. Inset shows system size scaling ($\bar{R}\propto N^{\alpha_{\bar{R}}}$) where $\alpha_{\bar{R}}\approx \gamma - 1$ for $\gamma\leq 1$ and 0 for $\gamma>1$.
	}
	\label{fig_10}
\end{figure}
In general, the survival probability decays till $t = \tTh$, known as the Thouless time \cite{Thouless3}. This is the time required for $\ket{j}$ to maximally spread over the Hilbert space. For example, in disordered (ergodic) metals, a particle diffuses to the sample boundaries within $\tTh$. The inverse of $\tTh$ gives the Thouless energy, $E_\text{Th}$ below which the spectral correlations are similar to those of Wigner-Dyson ensemble.
Moreover, a finite-sized closed quantum system always equilibrates \cite{Short1} and the equilibrium value of survival probability is given by
\begin{align}
	\label{eq_R_bar}
	\bar{R} = \lim\limits_{t\to\infty}\dfrac{1}{t}\int_{0}^{t}d\tau\;R(\tau) = \sum_{k=1}^N \abs{\phi_k^{(j)}}^4.
\end{align}
Thus $\bar{R}$ is the IPR of initial state $\ket{j}$ in the eigenbasis $\{\ket{\phi_k} \}$. The time required to reach $\bar{R}$ is known as the relaxation time, $\tR$. The gap between $\tTh$ and $\tR$ is known as the correlation hole, $t_\text{hole}$. A finite $t_\text{hole}$ is a direct manifestation of the spectral rigidity, i.e. the presence of long range correlation among energy levels \cite{Torres4,Torres2,Nosaka1}.

The time evolution of survival probability for \bte\ is shown in Fig.~\ref{fig_10}(a) and (b) for various system sizes and $\gamma$ values. Tuning $\gamma$, we observe three qualitatively different behaviours as follows:

{\it 1. Ergodic regime $(\gamma\leq 0):$} The correlation hole is always present  with easily identifiable Thouless and relaxation times. $\tTh$ exhibits an approximately $\sqrt{N}$ scaling close to $\get$ (inset of Fig.~\ref{fig_10}(a)), which can be understood from sparsity of the Hamiltonian. Contrarily in the ergodic regime of RPE, $\tTh$ is independent of system size due to the presence of all to all coupling \cite{Kravtsov1, Tomasi1}. We also observe that within $t_\text{hole}$, $R(t)$ is non-monotonic unlike \xxx\ \cite{Schiulaz1}.

{\it 2. NEE phase $(0<\gamma<1):$} We show $t_\text{hole}$ as a function of $\gamma$ for different system sizes in Fig.~\ref{fig_10}(c) where $t_\text{hole}\approx 0$ for $\gamma\geq \gamma^\star$. The inset shows that $\gamma^\star \rightarrow \get = 0$ as $N$ increases. Recall that chaotic-integrable transition occurs at $\get$, beyond which long range correlation among energy levels (e.g. see power spectrum) is lost. 
The spectral rigidity is necessary for the existence of $t_\text{hole}$ \cite{Torres4}, which explains the absence of 
correlation hole in the NEE regime in the thermodynamic limit. On the other hand, a finite $t_\text{hole}$ exists in the NEE phases of RPE and \xxx, as chaotic-integrable transition occurs at $\gtat$ and $h_\text{MBL}$ respectively.

{\it 3. Localized phase $(\gamma\geq 1):$} Exactly at $\gamma = \gat$, we observe survival probability curves from different system sizes to collapse on top of each other, showing a critical behaviour similar to RPE. In the localized regime (i.e. $\gamma>1$), $t_\text{hole}$ is completely absent while $R(t)$ converges to 1 upon increasing either $\gamma$ or $N$.

We show the equilibrium value of survival probability, $\bar{R}$, as a function of $\gamma$ in Fig.~\ref{fig_10}(d).
We observe that $\bar{R}\approx 1$ for $\gamma>1$ which is expected as the initial state, $\ket{j}$ is an eigenstate in the localized regime. Inset of Fig.~\ref{fig_10}(d) shows system size scaling of $\bar{R}$, indicating that $\bar{R}\approx N^{\gamma - 1}$ in the NEE regime denoting the extent of spread over the Hilbert space for an initially localized state.

\section{Conclusions}
In this work we study the spectral properties of \bte\ with a motivation that the competition between diagonal and off-diagonal terms may lead to a NEE phase. As customary in random matrix theory, we discuss the DOS and short-range spectral correlations, namely, NNS, RNNS and observe a transition from chaos to integrability at $\gamma = 0$. The next pertinent question is whether this transition can be associated with the ergodic and/or localization transitions. A simple analysis of the power spectrum of noise in the eigensequence identifies two critical points: ergodic transition at $\get = 0$ and Anderson transition at $\gat = 1$, separating three distinct phases: ergodic ($\gamma\leq 0$), NEE ($0<\gamma<1$) and localized ($\gamma\geq 1$) phase. Thus similar to RPE \cite{Kravtsov1}, related ensembles \cite{Biroli2, Khaymovich1} and certain floquet systems \cite{Ray1, Roy1, Sarkar1}, \bte\ is another matrix model where NEE phase exists over a finite interval of system parameters.

The above observations can be consolidated from the eigenfunction properties as both Shannon entropy and IPR show criticality at $\gat=1$, confirming it to be the Anderson transition point. The system size scaling of the above quantities gives us the fractal dimensions $D_{1, 2}\approx 1-\gamma$, clearly demarcating the three phases. For Relative R\'enyi entropies of type 1 and 2, criticality is seen at $\get=0$, thus confirming it to be the chaotic-integrable as well as the ergodic transition point.
Moreover, the distribution of Shannon entropy indicates that in the NEE phase, there is a coexistence of $N^\gamma$ number of completely localized and $(N - N^\gamma)$ number of NEE states.

Finally, we identify the relevant dynamical timescales from the time evolution of the survival probability, $R(t)$ of an initially localized state and find that the correlation hole, $t_\text{hole}$, is always present in the ergodic regime and absent in the NEE phase for $N\gg 1$ as energy levels become uncorrelated. Moreover $R(t)\to 1$ for $N\gg 1$ and $\gamma>\gat$ as expected in a localized phase. The NEE phase in \bte\ is quite distinct from that in RPE, where the energy levels repel each other since integrability breaks down at the Anderson transition point. Again the chaotic-integrable transition point is energy density dependent and ergodicity breaks at a lower disorder strength in the case of \xxx.
We calculate the entropic localization length to explain why chaotic-integrable transition in \bte\ does not coincide with the localization transition point. These subtle differences imply that \bte\ is not a suitable model for spin systems like \xxx. The proposition that \bte\ can model Heisenberg chain \cite{Buijsman1} has also been contradicted in \cite{Sierant1, Sierant2} via analyses of higher order level spacings, Spectral Form Factor (SFF) and in \cite{Corps1} by studying RNNS crossover. 
However, we observe that in the ergodic regime of the \bte, Thouless time scales with the system size similar to sparse Hamiltonians \cite{Schiulaz1}. Thus our analyses suggest that the \bte\ can imitate the spectral properties of various Hamiltonians provided ergodicity breaks down at the chaotic-integrable transition point in such systems.

{\bf Acknowledgment:} We thank Lea F. Santos and Ivan Khaymovich for many useful comments on the manuscript. AKD is supported by INSPIRE Fellowship, DST, India.
\appendix
\renewcommand\thefigure{\thesection.\arabic{figure}}  
\setcounter{figure}{0}
\renewcommand\thetable{\thesection.\arabic{table}}  
\setcounter{table}{0}
\section{Numerical details}\label{apnd_algo}
\paragraph{Unfolding:}For fixed values of system parameters and size, we obtain $\mathcal{F}(E)$, the cumulative density of eigenvalues from all  disordered realizations. Next we smooth $\mathcal{F}(E)$ using a moving average filter. Then for the original eigenvalue $E_i$, unfolding implies the interpolation of $\mathcal{F}(E_i)$ \cite{Guhr1}.
\paragraph{Scaling of crossover curves:}Let us look at an observable $y$ for a system parameter $x$. If we observe non-analytical behavior of the corresponding crossover curves from different system sizes, we assume $y$ to behave according to Eq.~\eqref{eq_power_scaling}. So we take an array of parameters $\vec{x} = (x(1),x(2),\dots,x(m))$ and measure $y$ for two system sizes $N_1$ and $N_2$, giving us two arrays $\vec{y_i} = (y_i(1), y_i(2),\dots, y_i(m)),\: i = 1,2$. Following Eq.~\eqref{eq_power_scaling}, we take the function $g(x; x_c, \nu, N)\equiv (x-x_c)(\log N)^{1/\nu}$. Now define the function $\vec{x_i}\equiv \vec{x_i}(x_c, \nu) = g(\vec{x}; x_c, \nu, N_i)$ for $i = 1, 2$. For the correct guess of $x_c$ and $\nu$, $\vec{y_1}$ vs. $\vec{x_1}$ and $\vec{y_2}$ vs. $\vec{x_2}$ should behave similarly. The observed non-analyticity of the curves $\vec{y_1}$ vs. $\gamma$ and $\vec{y_2}$ vs. $\gamma$ gives a good idea of the initial value of $x_c$ and we take $\nu = 1$ as the starting value. For these initial guesses of $x_c$ and $\nu$, we interpolate $\vec{y_1}$ vs. $\vec{x_1}$ w.r.t. $\vec{x_2}$ to create a new array $\vec{y_1'}$. Next we calculate the Residual Sum of Squares (RSS) between $\vec{y_1'}$ and $\vec{y_2}$, defined as $\sum_{k}\del{y_1'(k) - y_2(k)}^2$. Since RSS should be 0 for the correct choice of $x_c$ and $\nu$, we iteratively change $x_c$ and $\nu$ to minimize the RSS, which gives us the desired values of $x_c$ and $\nu$ for $N_1$ and $N_2$. We repeat this exercise for all pairs of system sizes and report the average of the obtained parameters.

\bibliographystyle{ieeetr}

\end{document}